\documentclass[epj,twocolumn]{webofc}
\usepackage[varg]{txfonts}   % Web of Conferences font
\usepackage{subfig}
\usepackage{subfig}
\usepackage{amsmath}
\usepackage{amssymb}
\usepackage{verbatim}
\usepackage{array}
\usepackage{times}

\newcommand{\be}{\begin{equation}}
\newcommand{\beq}{\begin{equation}}
\newcommand{\ee}{\end{equation}}
\newcommand{\eeq}{\end{equation}}
\newcommand{\eea}{\end{eqnarray}}
\newcommand{\bea}{\begin{eqnarray}}

\newcommand{\m}{\mathrm}

\woctitle{The Innermost Regions of Relativistic Jets and Their Magnetic Fields}
\begin{document}
\title{Uncovering the physics behind the blazar sequence using a realistic model for jet emission}

\author{William J. Potter\inst{1}\fnsep\thanks{\email{will.potter@astro.ox.ac.uk}} \and
        Garret Cotter\inst{1}
}

\institute{Oxford Astrophysics. Denys Wilkinson Building, Keble Road, Oxford, OX1 3RH. }

\abstract{%
Blazar spectra are one of the most important windows into the physical processes occurring along jets. The spectrum, composed from the different emitting regions along the jet, allows us to constrain the physical conditions in the jet. I present my work modelling blazar spectra using an extended inhomogeneous jet model with an accelerating, magnetically dominated, parabolic base transitioning to a slowly decelerating, conical section motivated by observations, simulations and theory. We set the inner geometry of our multi-zone model using observations of the jet in M87 which transitions from parabolic to conical at $10^{5}$ Schwarzschild radii.  This model is able to reproduce quiescent blazar spectra very well across all wavelengths (including radio observations) for a sample of 42 BL Lacs and FSRQs. 

Using this inhomogeneous model we are able to constrain the location at which the synchrotron emission is brightest in these jets by fitting to the optically thick to thin synchrotron break. We find that the radius of the jet at which the synchrotron emission is brightest (where the jet first approaches equipartition) scales approximately linearly with the jet power. We also find a correlation between the length of the accelerating, parabolic section of the jet and the maximum bulk Lorentz factor. In agreement with previous work we find that BL Lacs are low power blazars whereas FSRQs are high power blazars. Together with our simple jet power-radius relation this leads us to a deeper understanding of the physics underlying the blazar sequence.
}
\maketitle
\section{Introduction}
\label{intro}

It has recently been observed that the jet in M87 has a parabolic base out to $10^{5}$ Schwarzschild radii, becoming conical beyond this distance \cite{2011arXiv1110.1793A}. This is consistent with state of the art numerical simulations which show the jet base starts magnetically dominated and accelerates in the parabolic region reaching a terminal bulk Lorentz factor at equipartition, after which the jet becomes ballistic and conical \cite{2006MNRAS.368.1561M}, \cite{2009MNRAS.394.1182K}, \cite{2010ApJ...711...50T} and \cite{2012MNRAS.423.3083M}. Here we use an emission model with a magnetically dominated, accelerating, parabolic base transitioning at equipartition to a slowly decelerating conical jet, to test this emerging picture of jets and to try to determine their properties. The model (developed in \cite{2012MNRAS.423..756P}, \cite{2013MNRAS.429.1189P}, \cite{2013MNRAS.431.1840P} and \cite{2013MNRASP4.}) is used to fit to all 38 blazars with simultaneous multiwavelength observations and redshifts from the {\it Fermi} sample \cite{2010ApJ...716...30A} and the 4 quasi-simultaneous HSP BL Lacs from \cite{2012MNRAS.422L..48P}.

\begin{figure}

		 \includegraphics[width=11cm, clip=true, trim=4cm 1cm 1cm 1cm]{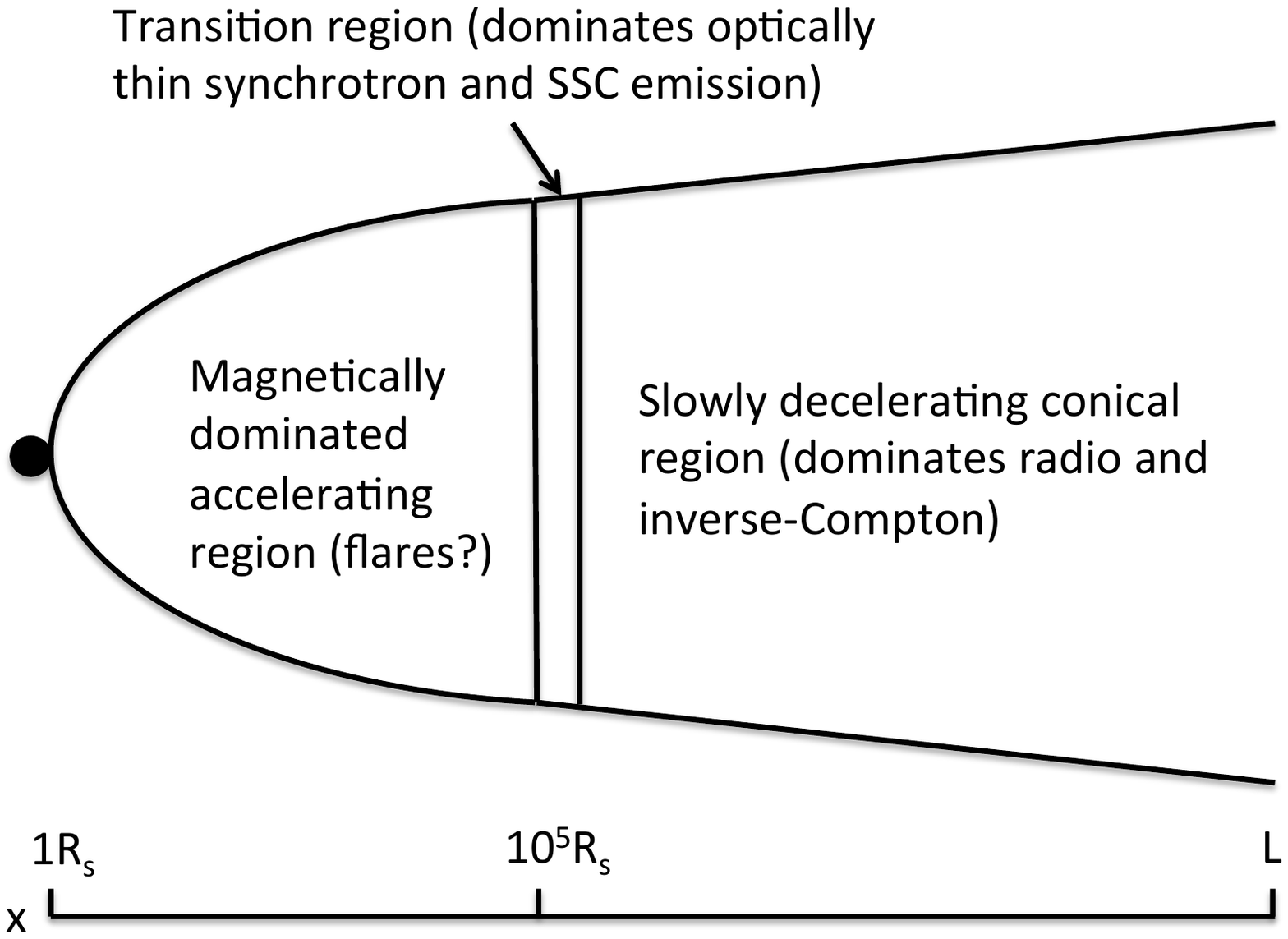}

	\caption{A schematic diagram of our jet model (taken from \cite{2013MNRAS.431.1840P}).}
	\label{schematic}
\end{figure}

In this model (see Figure \ref{schematic}) we split the parabolic to conical jet geometry into cylindrical sections ($\sim500$) with adaptively chosen widths. Relativistic energy-momentum and particle number are conserved along the jet. Magnetic energy is converted into bulk kinetic energy in the accelerating, magnetically dominated region and bulk kinetic energy is converted into in-situ particle acceleration in the decelerating conical region. In the conical section of the jet electron radiative losses are compensated by in situ particle acceleration (with an initial electron distribution in the form of a power-law with an exponential cutoff) in order for the plasma to remain close to equipartition. Both radiative energy losses from synchrotron and inverse-Compton emission and adiabatic energy losses to the electron distribution are calculated. We include a detailed treatment of all the relevant sources of external photons as a function of distance along the jet including accretion disc, BLR, dusty torus, NLR, starlight and CMB photons. These external photon fields are Lorentz transformed into the rest frame of the plasma at each section of the jet and the inverse-Compton emission is calculated using the full Klein-Nishina cross-section. For full details of our jet model see \cite{2012MNRAS.423..756P} and \cite{2013MNRAS.429.1189P}.

\begin{figure*}
	\centering
		\subfloat[]{ \includegraphics[width=12cm, clip=true, trim=2cm 1cm 1cm 2cm]{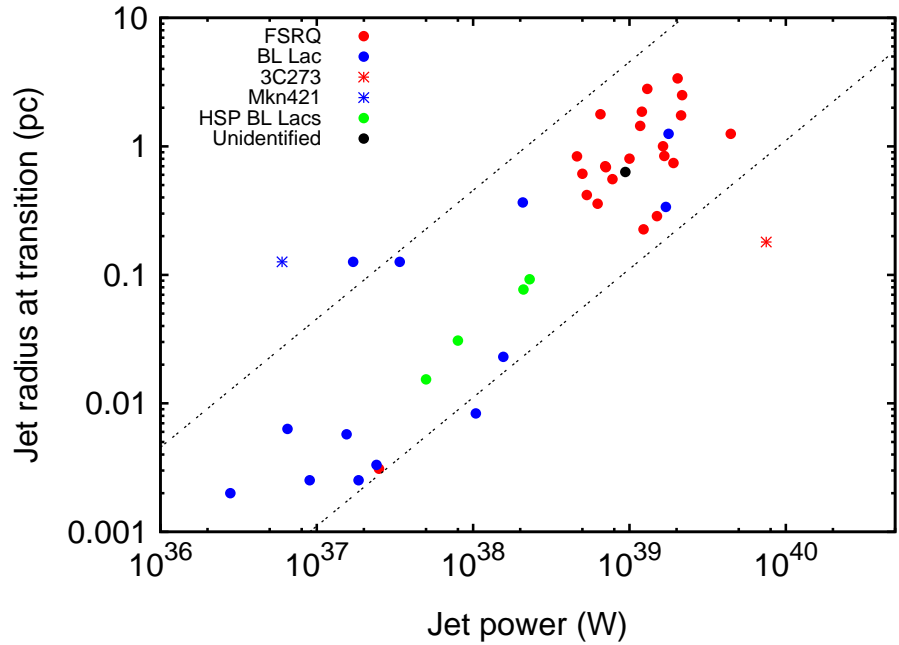} }
\\
		\subfloat[]{ \includegraphics[width=12cm, clip=true, trim=2cm 1cm 1cm 2cm]{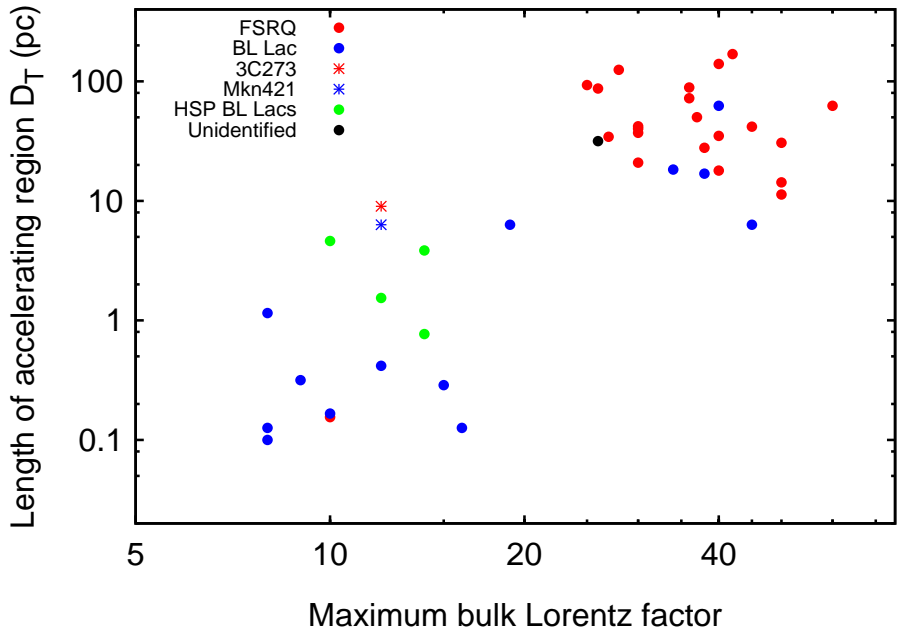} }	
		
	\caption{Fits to all 38 blazars with simultaneous multiwavelength spectra and redshifts and the 4 quasi-simultaneous HSP BL Lac SEDs. Figure a: the distance of the jet transition region plotted against the jet power for all fits. There is a clear correlation between the jet power and radius of the transition region as suggested in our previous work. The correlation between the jet power and transition region radius is approximately linear as illustrated by the dashed lines. BL Lacs have systematically smaller radius transition regions than FSRQs. We have highlighted the two outliers to the relation, Markarian 421 and 3C273. We argue that these two blazars are atypical since Mkn421 is one of the closest known blazars and 3C273 is the one of the most powerful (and seems to be misaligned). Figure b: The maximum bulk Lorentz factor plotted against the length of the accelerating region of the jet. The bulk Lorentz factor is systematically larger for FSRQs than BL Lacs and increases with the distance over which the jet accelerates as we expect. }
	\label{corr}
\end{figure*}

\begin{figure*}
	\centering
		\subfloat[]{ \includegraphics[height=5.5cm, clip=true, trim=2cm 1cm 1cm 2cm]{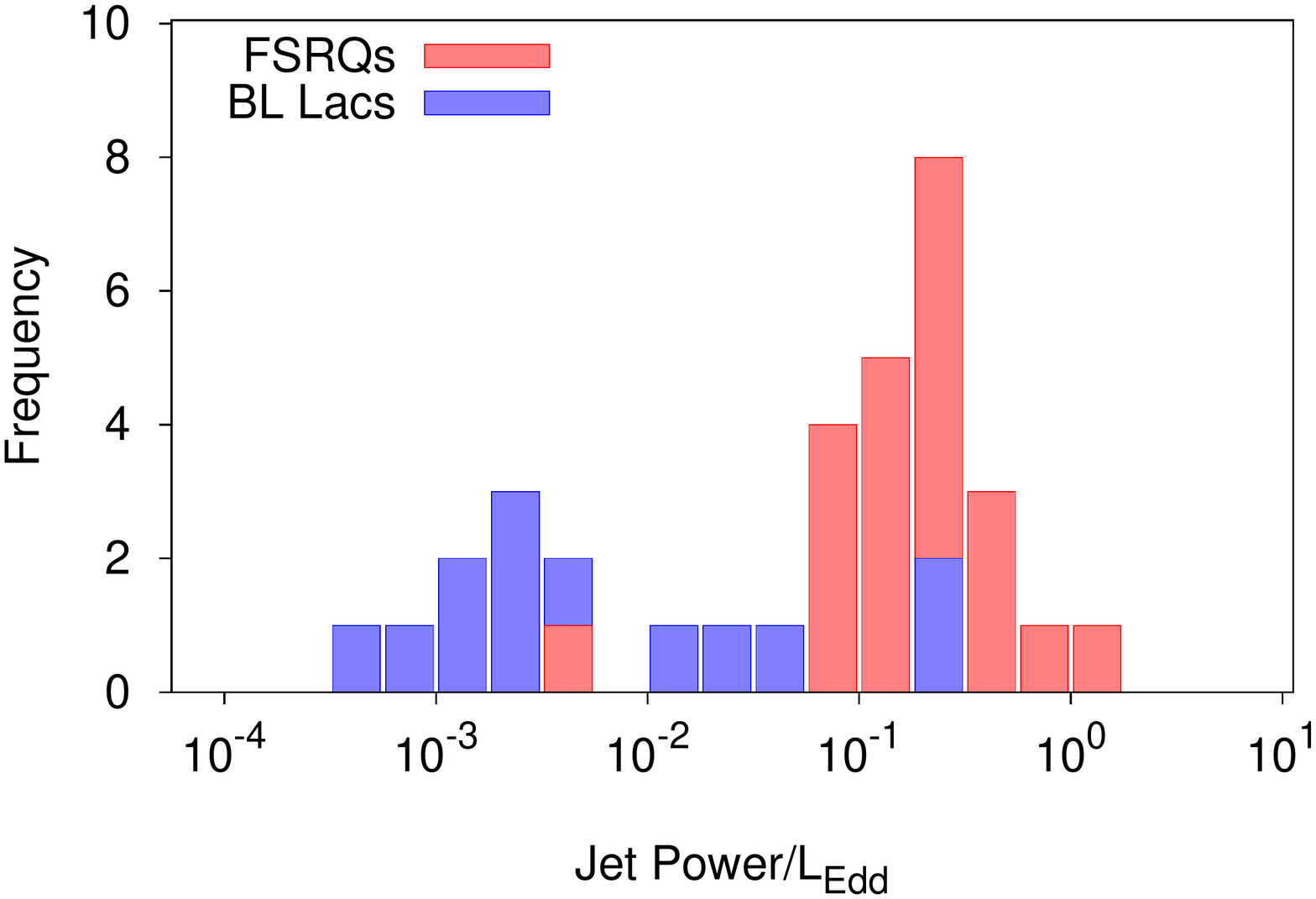} }
		\subfloat[]{ \includegraphics[height=5.5cm, clip=true,  trim=2cm 1cm 1cm 2cm]{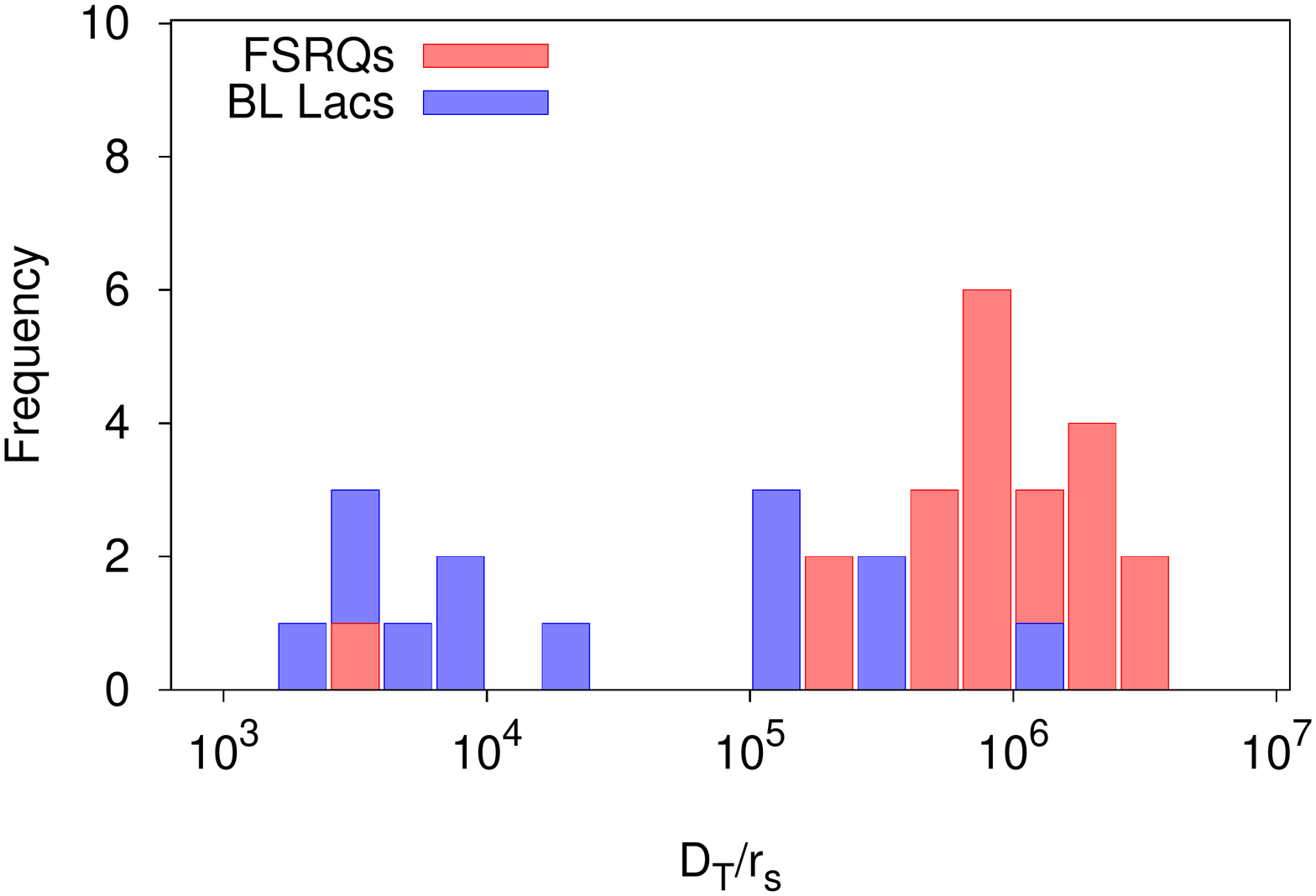} }	
	\caption{The distribution of fractional Eddington luminosity and the distance to the jet transition region under the assumption of a constant black hole mass of $5\times 10^{8}M_{\odot}$ for all 38 blazars with simultaneous observations. Figure a: A clear dichotomy in the fractional Eddington luminosity between BL Lacs and FSRQs, suggestive of two accretion modes. Figure b: The accelerating transition region extends out to larger distances in FSRQs than BL Lacs, this could be responsible for the different maximum bulk Lorentz factors shown in Figure \ref{corr}b. }
	\label{distributions}
\end{figure*}

\begin{figure*}
	\centering
		\subfloat{ \includegraphics[width=10cm, clip=true, trim=2cm 1cm 1cm 1cm]{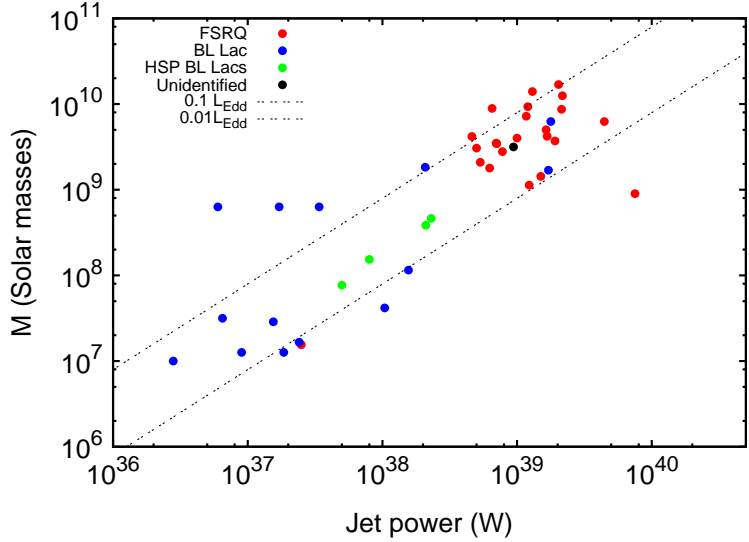} }
		
	\caption{ The inferred black hole mass of the fits plotted against jet power assuming the transition region occurs at a distance of $10^{5}r_{s}$ as in M87. The range of black hole masses is compatible with that found from large galaxy surveys ($10^{7}$--$10^{10}M_{\odot}$). A linear relation between jet power and the radius of the transition region is found naturally if we assume that all blazars are accreting at a similar Eddington fraction $0.01-0.1L_{Edd}$.}
	\label{corr3}
\end{figure*}

\begin{figure*}
	\centering
		\subfloat{ \includegraphics[height=5.5cm, clip=true, trim=1cm 1cm 0cm 1cm]{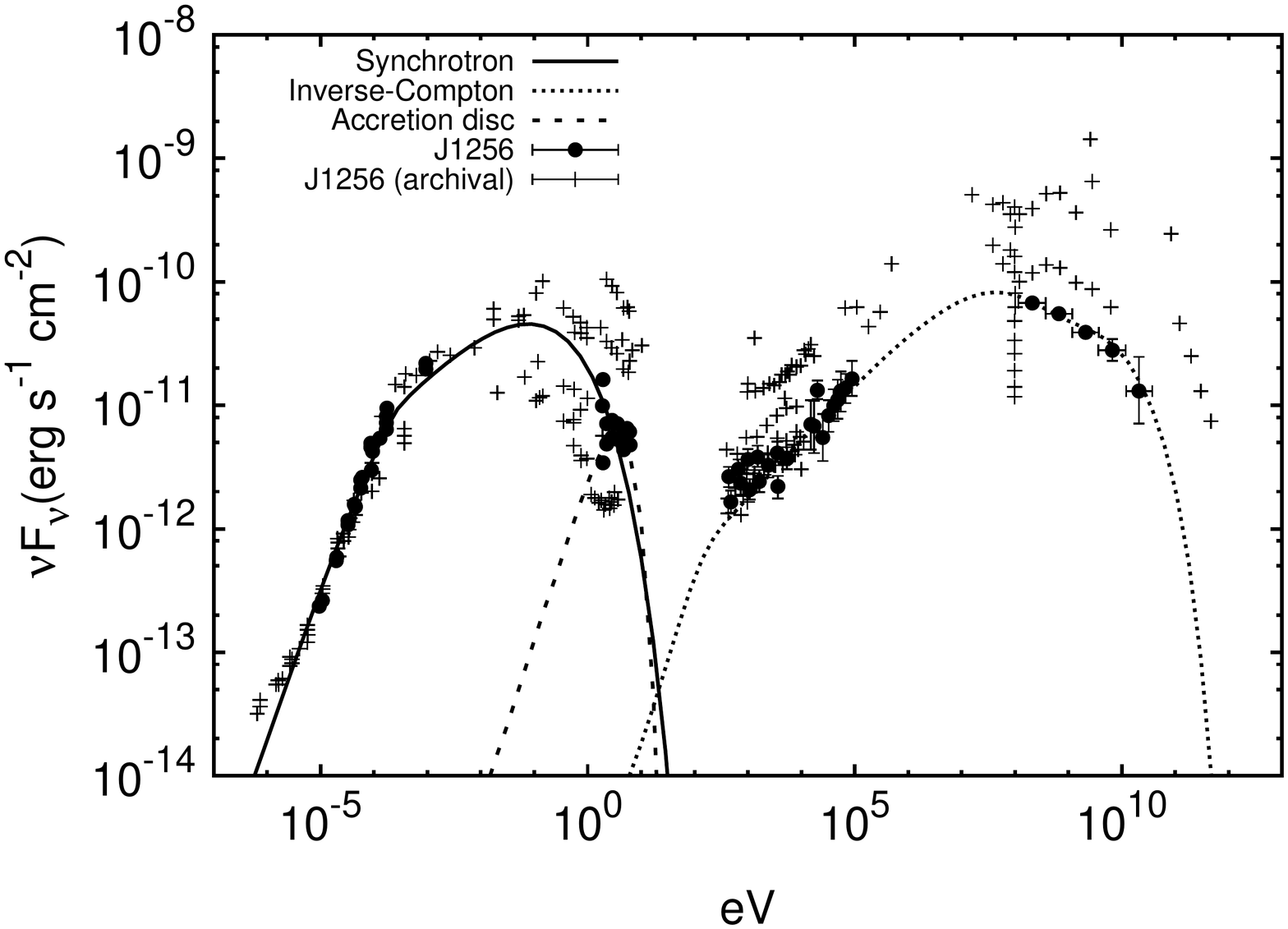} }
		\subfloat{ \includegraphics[height=5.5cm, clip=true,  trim=1cm 1cm 0cm 1cm]{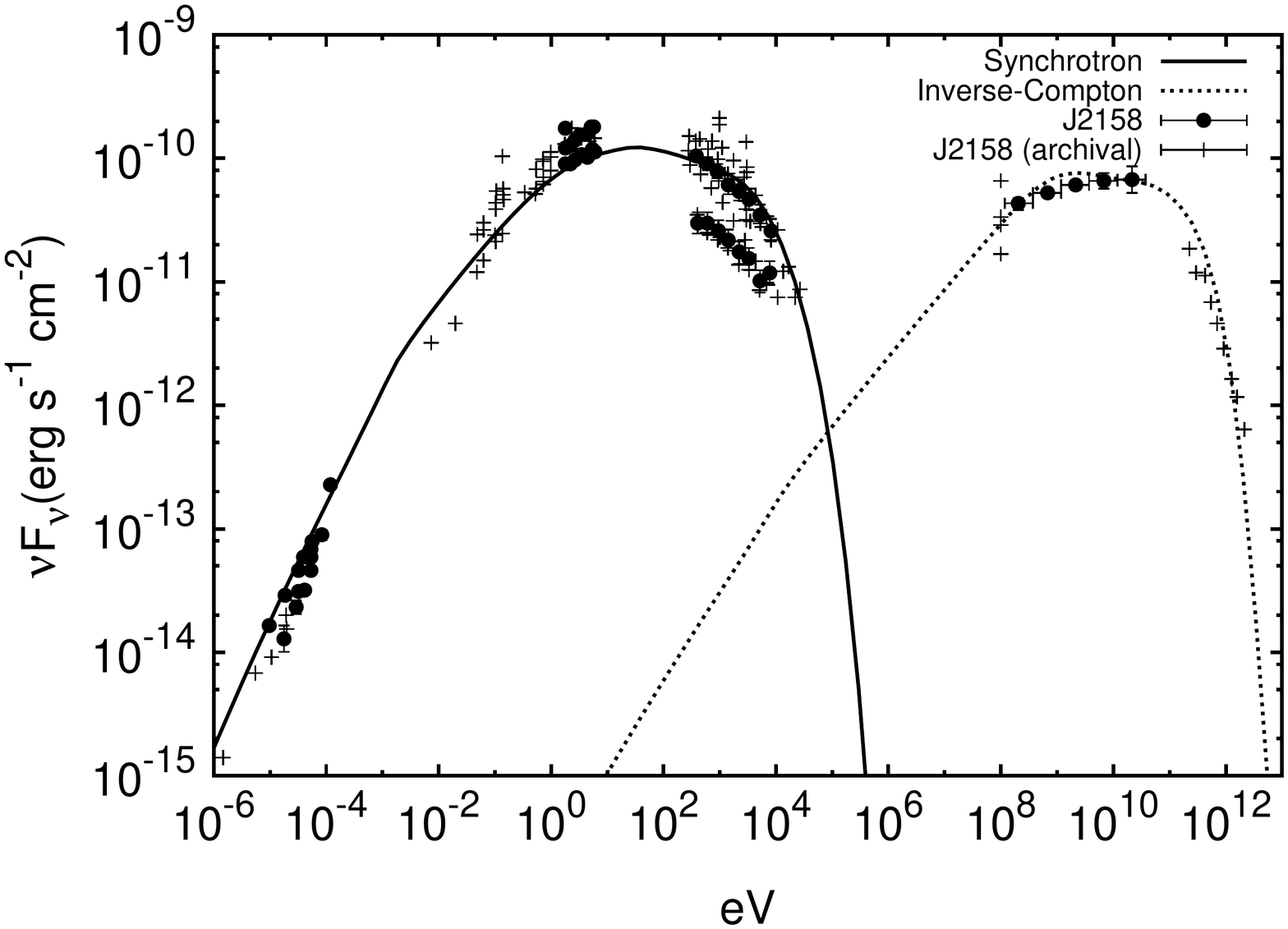} }	
	\caption{Example model fits to J2158.8$-$3014 (BL Lac) and J1256.1$-$0547 (FSRQ). Our model is able to fit to the observations of all 42 blazars extremely well across all frequencies, including radio data.  }
	\label{examples}
\end{figure*}

\section{Results}
\label{Results}

The main results of fitting our model to the 42 blazars are shown in Figure \ref{corr}. There is a clear correlation between the jet power and transition region radius (where the jet first comes into equipartition and transitions from parabolic to conical). The correlation is approximately linear as illustrated by the dashed lines. Figure \ref{corr}b shows a correlation between the maximum bulk Lorentz factor and the length of the accelerating region, $D_{T}$, where $D_{T}=50R_{T}$ set from M87 and $R_{T}$ is the transition region radius. There is a clear separation between FSRQs and BL Lacs in jet power as we expect from previous work. High power FSRQs have larger radius transition regions and larger maximum bulk Lorentz factors than BL Lacs. These properties offer support for the unification of FSRQ and BL Lacs with FRII and FRI jets respectively.

This is the first inhomogeneous jet model which has been able to successfully reproduce blazar observations across all frequencies including the flat, self-absorbed radio spectrum produced by the extended large scale structure of the jet. By fitting to the radio observations we gain important additional constraints on the jet structure, which have been previously neglected. The most significant of these constraints comes from fitting to the optically thick to thin synchrotron break at low frequencies. In our previous work \cite{2013MNRAS.431.1840P} the synchrotron break frequency was shown to depend sensitively on the radius of the jet which dominates the optically thin jet emission (in our model this is the transition region where the jet first comes into equipartition). The synchrotron break can be used to constrain the radius of the transition region from fitting to the SED. In order to fit to the synchrotron break FSRQs require the transition region to occur at a distance $>10$pc. This places the region of the jet which first comes into equipartition and dominates the optically thin synchrotron emission outside of the BLR and dusty torus. This is significant because it is usually assumed that the synchrotron emitting region is located within the BLR or dusty torus. At these large distances outside the BLR and dusty torus we find that the gamma-ray emission is very well fitted by scattering CMB photons at large distances along the jet, with a small contribution due to the scattering of NLR and starlight photons. The slope of the radio spectrum is systematically steeper in FSRQs than BL Lacs. The radio spectrum of BL Lacs are fitted well by a ballistic conical jet in equipartition, whilst FSRQs require a jet which starts in equipartition at the transition region and becomes particle dominated at large distances. 

The results in Figure \ref{corr} allow us to understand the physics behind the blazar sequence. Low power jets with smaller transition region radii have larger magnetic field strengths at the synchrotron bright transition region and so have high peak synchrotron frequencies. Their inverse-Compton emission is due predominantly to SSC because of their high magnetic fields and compact emitting regions, so the number density of synchrotron photons is large. They are not Compton-dominant because the \lq\lq{}Compton-catastrophe\rq\rq{} prevents SSC emitting regions from becoming Compton-dominant easily. High power blazars have larger transition regions with lower magnetic field strengths resulting in lower synchrotron peak frequencies and sub-dominant SSC emission. Their inverse-Compton emission is best fitted by scattering CMB and narrow line region (NLR) photons at large distances if their jets remain relativistic up to kiloparsec scales. These properties result in the blazar sequence: the observed anticorrelation between peak emitting frequencies and jet power, and Compton-dominance and jet power.

We can interpret the results in Figure \ref{corr} through two simple scenarios depending on the black hole masses of BL Lacs and FSRQs. The first scenario is that BL Lacs and FSRQs have similar black hole masses as indicated by black hole mass estimates (see \cite{2012ApJ...748...49S} and \cite{2013ApJ...764..135S}). Using a fiducial value of $5\times 10^{8}M_{\odot}$ for simplicity, the distribution of Eddington luminosity and length of the accelerating region of the jet in BL Lacs and FSRQs are shown in Figure \ref{distributions}. In this scenario we find an approximately bimodal distribution of Eddington luminosity in BL Lacs ($0.001-0.1L_{\m{Edd}}$) and FSRQs ($0.1-1L_{\m{Edd}}$), analogous to the different accretion modes in X-ray binary systems. The length of the accelerating region is larger for FSRQs than BL Lacs (Figure \ref{distributions}b). This is to be expected since jets which accelerate over larger distances (measured in Schwarzschild radii) should have larger bulk Lorentz factors \cite{2003ApJ...596.1080V}, consistent with Figure \ref{corr}b.

The second scenario we consider is that of a universal jet geometry in which all jets have the same inner structure as M87 with a transition region occurring at $10^{5}r_{s}$. In this case the inner jet structure scales linearly with black hole mass. The inferred black hole masses of the blazar fits in this scenario are shown in Figure \ref{corr3}. The range of black hole masses is plausible, $10^{7}-10^{10}M_{\odot}$, with BL Lacs having lower masses than FSRQs. A linear relation between transition region radius and jet power naturally arises in this scenario if all blazars are accreting at a similar Eddington fraction $0.01-0.1L_{\m{Edd}}$. Currently, we disfavour this second interpretation of our results due to the inconsistency with black hole mass estimates.

Our extended jet model is able to fit very well to the multiwavelength spectra of all 42 blazars across all frequencies including radio data (example fits are shown in Figure \ref{examples}).

\section{Conclusion}
\label{Conclusion}

We have shown that our jet model with a magnetically dominated, accelerating, parabolic base transitioning to a slowly decelerating, conical jet is able to fit to observations of blazar jets with unprecedented detail across all frequencies. This offers support for the emerging picture of jets motivated by observations and simulations. Fitting to a large sample of 38 simultaneous, multiwavelength blazar spectra we are able to constrain the dynamic and structural properties of blazar jets. Our two main results are: the discovery of a linear correlation between the radius of the transition region of the jet and the jet power, and a relation between the maximum bulk Lorentz factor and the distance over which the jet accelerates before reaching equipartition. These results allow us to understand the physical origin of the blazar sequence. We also find evidence for an accretion mode dichotomy in AGN analogous to that observed in X-ray binary systems and consistent with a correspondence between BL Lacs and FRI-type jets, and FSRQs and FRII-type jets.

\bibliographystyle{woc}
\bibliography{jetpaper2refs}

\end{document}